\documentclass[journal,twoside,web]{IEEEtran}
\usepackage{amsmath,amssymb,amsfonts}
\usepackage[ruled,linesnumbered]{algorithm2e}
\usepackage{graphicx}
\usepackage{hyperref}
\usepackage[flushleft]{threeparttable}
\usepackage{multirow}
\usepackage{float}
\usepackage{subcaption}
\usepackage{soul}
\usepackage{changes}
\hypersetup{hidelinks=true}
\usepackage{textcomp}
\begin{document}
\title{ECG Biometric Authentication Using Self-Supervised Learning for IoT Edge Sensors}
\author{Guoxin Wang, \IEEEmembership{Member, IEEE}, Shreejith Shanker, \IEEEmembership{Member, IEEE}, Avishek Nag, \IEEEmembership{Senior Member, IEEE}, Yong Lian, \IEEEmembership{Fellow, IEEE}, and Deepu John \IEEEmembership{Senior Member, IEEE}
\thanks{This work is partly supported by 1) the China Scholarship Council, 2) the JEDAI Project under the Horizon 2020 FET Chist-Era Program and 3) Microelectronic Circuit Centre Ireland.}
\thanks{Guoxin Wang and Deepu John are with the School of Electrical and Electronic Engineering, University College Dublin, Dublin 4, Ireland (e-mail: guoxin.wang@ucdconnect.ie; deepu.john@ucd.ie).}
\thanks{Avishek Nag is with the School of Computer Science, University College Dublin, Dublin 4, Ireland (e-mail: avishek.nag@ucd.ie).}
\thanks{Yong Lian is with the Dept of EECS, York University, Toronto, Canada (e-mail: plian@yorku.ca).}
\thanks{Shreejith Shanker is with the Department of Electronic and Electrical Engineering, Trinity College Dublin, Dublin 2, Ireland (e-mail: shankers@tcd.ie).}}
\maketitle

\begin{abstract}
Wearable Internet of Things (IoT) devices are gaining ground for continuous physiological data acquisition and health monitoring. 
These physiological signals can be used for security applications to achieve continuous authentication and user convenience due to passive data acquisition.
This paper investigates an electrocardiogram (ECG) based biometric user authentication system using features derived from the Convolutional Neural Network (CNN) and self-supervised contrastive learning. Contrastive learning enables us to use large unlabeled datasets to train the model and establish its generalizability. We propose approaches enabling the CNN encoder to extract appropriate features that distinguish the user from other subjects. 
{When evaluated using the PTB ECG database with 290 subjects, the proposed technique achieved an authentication accuracy of 99.15\%. To test its generalizability, we applied the model to two new datasets, the MIT-BIH Arrhythmia Database and the ECG-ID Database, achieving over 98.5\% accuracy without any modifications. Furthermore, we show that repeating the authentication step three times can increase accuracy to nearly 100\% for both PTBDB and ECGIDDB.}
This paper also presents model optimizations for embedded device deployment, which makes the system more relevant to real-world scenarios. 
{To deploy our model in IoT edge sensors, we optimized the model complexity by applying quantization and pruning. The optimized model achieves 98.67\% accuracy on PTBDB, with 0.48\% accuracy loss and 62.6\% CPU cycles compared to the unoptimized model.}
An accuracy-vs-time-complexity tradeoff analysis is performed, and results are presented for different optimization levels.
\end{abstract}


\section{Introduction}
Biometric authentication is widely accepted as the security mechanism in many cyber-physical system applications.
Typically, biological or behavioural characteristics such as fingerprints, voice snippets, face, and iris scanning, among others, are used at the start of a session to authenticate a user and determine their access privileges \cite{zheng2018finger}. Once authenticated, the session remains active until a period of inactivity has elapsed or till the user logs out. 
Additionally, techniques like continuous user authentication can be integrated with biometric schemes to restrict unapproved handover of system controls during inactive periods \cite{smyth2021continuous}. 
The higher convenience and security biometric authentication offers over traditional schemes such as key cards and passwords enabled their widespread adoption.

{While various biometric authentication systems are in use, integrating continuous authentication without overhead or user disruption is challenging \cite{wazid2017novel}. Typically, systems require periodic re-authentication, affecting user experience. A promising alternative is leveraging personal wearable devices to collect biometric data like activity and positional information in the background without disrupting the user \cite{panayides2020ai}. The electrocardiogram (ECG) stands out for continuous authentication among wearable-acquired biometric signals due to its universality, distinctiveness, and resistance to replication \cite{sun2022perae}. Additionally, ECG data is more accessible for research than behavioural biometrics \cite{stylios2021behavioral}. Convolutional Neural Networks (CNNs) effectively extract unique identifiers from the time-series representation of ECG signals \cite{singh2022interpretation}, which can then be processed for user authentication using machine learning techniques.}

Existing ECG-based authentication systems use the same small database for training and testing. In addition, these works do not test on unseen datasets, which cannot show a credible generalizability \cite{tantawi2015wavelet, hammad2018multimodal, thentu2021ecg, prakash2022baed, sepahvand2021novel, hazratifard2023ensemble}.

Although relatively straightforward, ECG-based authentication using wearable sensors faces some unique challenges and/or competing requirements: 

\begin{itemize}
    \item Limited system scalability, as fully supervised learning approaches are used, and only a fixed number of users are present at training time. This means the system needs to be retrained to accommodate every additional user \cite{sarker2021machine}.
    \item The time-varying nature of ECG (e.g. based on user activity)  will result in signal variations for the same user. This requires a robust feature extraction step to capture unique and appropriate features.
    \item High-accuracy requirements for authentication, particularly in critical systems.
    \item Limited compute resources in wearable devices require lightweight and streamlined algorithms.
\end{itemize}

This paper proposes a novel framework for biometric authentication based on ECGs to be implemented in IoT wearable edge devices to solve these challenges. We propose to use contrastive learning, a self-supervised approach, to address the difficulties with system scaling. With our approach, the system  1) does not need fully annotated or labelled data for training and 2) does not require further model updating for a new unregistered user to prevent unauthorized access. We also propose using a CNN encoder to extract optimal features from ECG. Thus, it avoids using handcrafted features and increases the generalizability and accuracy when evaluated with previously unseen data by the model. Finally, we compute the Pearson Correlation Coefficient between registered users and current input to authenticate users. We also deploy the proposed algorithm in an ARM Cortex-M4F-based embedded microcontroller and report the accuracy and power consumption results. To reduce complexity, we apply model compression techniques such as quantization and pruning before the embedded edge implementation.     
Our experiments show that the proposed system outperforms the state-of-the-art (traditional supervised training with distance decision method from \cite{sepahvand2021novel}) authentication accuracy. The model achieves excellent results when tested with previously unseen datasets, demonstrating its robustness and generalizability. In addition, the contrastive learning training framework we designed has broader applications, i.e., it enables large-scale unsupervised pre-training for ECG-related tasks with multiple datasets that do not require pre-annotated data.

The contributions of this work are as follows:

\begin{itemize}
    \item A novel ECG authentication framework is proposed based on self-supervised contrastive learning. To the best of our knowledge, this is the pioneering work employing contrastive learning in the context of ECG authentication. Multiple data segmentation methods are evaluated. The proposed system achieves a high accuracy of 99.15\%.
    \item The proposed technique does not require handcrafted features followed by manual feature selection. The features are learned representations. In addition, our approach makes the independent labelling for ECG records unnecessary; therefore, the proposed system enables more accessible unsupervised model training.
    \item {The proposed model is trained using one large dataset and tested using previously unseen datasets with accuracies of 98.67\% and 98.77\%. The generalizability of the model is thus established.}
    \item We also optimized the final model using compression techniques to reduce the overall complexity. The performance trade-off analysis at various sparsity levels and power consumption, when implemented in an ARM Cortex M4F CPU-based SoC, is presented.
\end{itemize}

The rest of the paper is organized as follows: Section \ref{sec: rel} reviews related work in recent years. Section \ref{sec: proposed} introduces the system architecture and the authentication workflow, while the training framework is covered in Section \ref{sec: training}. Section \ref{sec: optim} provides high-accuracy optimization and low-complexity improvements. Section \ref{sec: results} presents the evaluation methodology and results from our experiments. Section \ref{sec: concl} concludes the paper, outlining possible extensions.

\section{Related Work}\label{sec: rel}
This section briefly reviews the state-of-the-art literature on ECG biometric authentication and the performance of CNN-based feature extraction for ECG-based user authentication applications.

\subsection{{ECG Feature Extraction}}
ECG biometric approaches can be divided into fiducial, non-fiducial, and hybrid methods based on the features used \cite{lin2020wearable}. ECG signals have several characteristic points (P, Q, R, S, T), and fiducial methods use these points to compute fiducial measurements such as amplitude, angle, and intervals. The non-fiducial methods extract the ECG features without characteristic point detection. Usually, these feature extractions are implemented using signal processing techniques. In hybrid methods, the characteristic points divide ECG into several segments, and signal processing techniques are used.

\subsubsection{{Fiducial Methods}}
{Tantawi et al. \cite{tantawi2015wavelet} extracted RR intervals and decomposed the intervals using discrete biorthogonal wavelets. Further, a radial-basis-function neural network was used for authentication, which achieved $97.7\%$ accuracy over $290$ subjects on PTBDB. This work needs handcrafted features. In addition, their method cannot detect unseen users. Yang et al. \cite{yang2021privacy} designed a scheme based on fiducial method feature extraction. This solution was deployed on a Raspberry Pi and achieved an error rate of $8.67\%$ on PTBDB. This work requires handcrafted features and reports a high error rate.}

\subsubsection{{Non-fiducial Methods}}
{In \cite{hejazi2016ecg}, Hejazi et al. built a non-fiducial ECG authentication system using Discrete Wavelet Transform (DWT) based feature extraction. A multi-class Support Vector Machine (SVM) with a one-against-all approach is used to authenticate users. This framework achieves $94.54\%$ accuracy when tested on the short-term rest ECG database from the University of Toronto with $52$ subjects. This method requires complex data preprocessing, and the normalised autocorrelation method does not extract all relevant features, which results in low accuracy. In addition, their experimental dataset was small, making the results unreliable. Agrafioti et al. \cite{agrafioti2010signal} presented an ECG-authentication method that extracted ECG features by Autocorrelation-Linear Discriminant Analysis and enhanced it by incorporating the periodicity transform. They used Euclidean distance to compute the similarity between different transformations, classify subjects using a k-nearest neighbour model, and reported an accuracy of $92.3\%$ over $52$ subjects on the short-term rest ECG database. They face the same challenges of complex processing and low performance.} 
{Huang et al. \cite{huang2019practical}'s proposed system achieved an F1-score of $97\%$ on a small collected test set, using an IoT device for signal collection and a laptop for data processing and authentication. Their work is only data collection with embedded devices and does not integrate model inference, which does not fit the requirements for decreasing complexity and integrating collection and inferencing.}

\subsubsection{{CNN-based Methods}}
CNNs are highly successful in image recognition tasks. 
CNN models can be trained such that the convolutional filters (encoder) can extract discriminatory features from raw data without knowledge of underlying feature composition signals. 
Therefore, the CNN encoder can be used for efficient feature extraction in ECG biometric authentication.
Several such techniques have been previously investigated. 
In \cite{hammad2018multimodal}, Hammad et al. used CNN to perform feature extraction and generated biometric templates from these features. They proposed a Q-Gaussian multi-support vector machine (QG-MSVM) as the classifier and reported an accuracy of $98.66\%$ over $290$ subjects on PTBDB. Labati et al. 
{In \cite{ghazarian2022assessing}, Ghazarian et al. explored using ECGs for human identification by training convolutional neural networks on data from approximately 81,000 patients, achieving 95.69\% accuracy. Their findings highlight the privacy risks of ECG data, as anonymized datasets can still lead to patient reidentification.}
\cite{labati2019deep} proposed 'Deep-ECG', which used CNN for feature extraction and Hamming distance for user authentication. Their system achieves $100\%$ accuracy over $52$ subjects on MITDB.

{The existing methods usually showed weak generalizability because they train and test on a single dataset \cite{tantawi2015wavelet, yang2021privacy, hejazi2016ecg, agrafioti2010signal, huang2019practical}, which easily overfits the model. With contrastive learning frameworks, we achieve high accuracy on unseen datasets, which shows strong generalizability. In addition, these high-computing-based approaches are often difficult to deploy practically due to high resource consumption \cite{hammad2018multimodal, ghazarian2022assessing, labati2019deep}. We deploy models on the embedded system to solve this challenge.}

\subsection{Contrastive Learning}
Most mainstream machine learning approaches are supervised methods, which rely on pre-annotated labels. This has some drawbacks:

\begin{itemize}
    \item Supervised methods do not spontaneously extract features from the data but rely excessively on labelled data. Training models using supervised methods require a large amount of labelled data, and the resulting models are sometimes fragile due to labelled errors.
    \item Supervised learning tends to learn only task- and dataset-specific knowledge, but not general knowledge, so the feature representations learned by supervised learning are complex to transfer to other tasks and datasets \cite{jaiswal2020survey}.
\end{itemize}

Self-supervised learning avoids these problems because it uses the data to provide label information to guide learning. Contrastive learning is a self-supervised approach that learns the signal features by comparing the data with positive and negative samples in the feature space, respectively \cite{le2020contrastive}. Contrastive learning does not focus on the tedious details of instances. Rather, it only distinguishes data on the feature space at the abstract semantic level, making the model and its optimization more straightforward and generalizable.

Several contrastive learning approaches have been reported in the literature. He et al. \cite{he2020momentum} proposed an efficient structure called MoCo for contrastive learning. The features learned using the MoCo-based unsupervised learning structure for ImageNet classification can exceed the performance of supervised learning. Chen et al. \cite{chen2020simple} designed the SimCLR framework, which performs data augmentation on the input images to simulate the input under different conditions. After that, a contrastive loss is used to maximize the similarity of the same targets under different data augmentation and minimize the similarity between similar targets.

Several frameworks leverage contrastive learning techniques in the context of ECG signal analysis. Chen et al. \cite{chen2021clecg} introduced an instance-level ECG pre-training framework, encouraging similar representations for different augmented views of the same signal and increasing the distance between representations from different signals. This strategy demonstrated improved performance in atrial fibrillation classification compared to conventional methods. Additionally, Wei et al. \cite{wei2022contrastive} proposed contrastive heartbeats, focusing on learning general and robust ECG representations for efficient linear classification training. Their approach utilizes a novel heartbeat sampling method to define positive and negative pairs for contrastive learning, leveraging the periodic and meaningful patterns in ECG signals.

Existing feature-based approaches usually cannot fully utilize the information in the ECG signal (i.e. lost information during transformation and feature selection), which may affect its accuracy. Further, since these frameworks pre-trained the classifier on a known dataset and subsequently employed the same classifier for one-vs-all identification, they usually cannot authenticate unknown, unregistered users, constraining their practical utility \cite{hejazi2016ecg, agrafioti2010signal, tantawi2015wavelet}.

Although some algorithms improve authentication accuracy by deploying a CNN feature extractor, they have only been tested on small datasets and do not demonstrate generalizability. In addition, these CNN feature extractors are obtained by training with an identification-oriented task, which results in the trained CNN networks being only available to a limited number of users in the authentication task \cite{hammad2018multimodal, labati2019deep}. Because they are not optimized for algorithmic complexity, they can only be deployed on platforms with high computational resources.

Contrastive learning is proven effective in learning feature differences and is well-suited for ECG-based authentication \cite{chen2021exploring, grill2020bootstrap}. In addition, deploying models on IoT devices requires reducing time and space complexity. Based on the above considerations, we proposed a CNN-based system that uses contrastive learning to train and optimize the model performance. 

\section{Proposed ECG Biometric System}\label{sec: proposed}
\begin{figure}[ht]
    \centering
    \includegraphics[width=\linewidth]{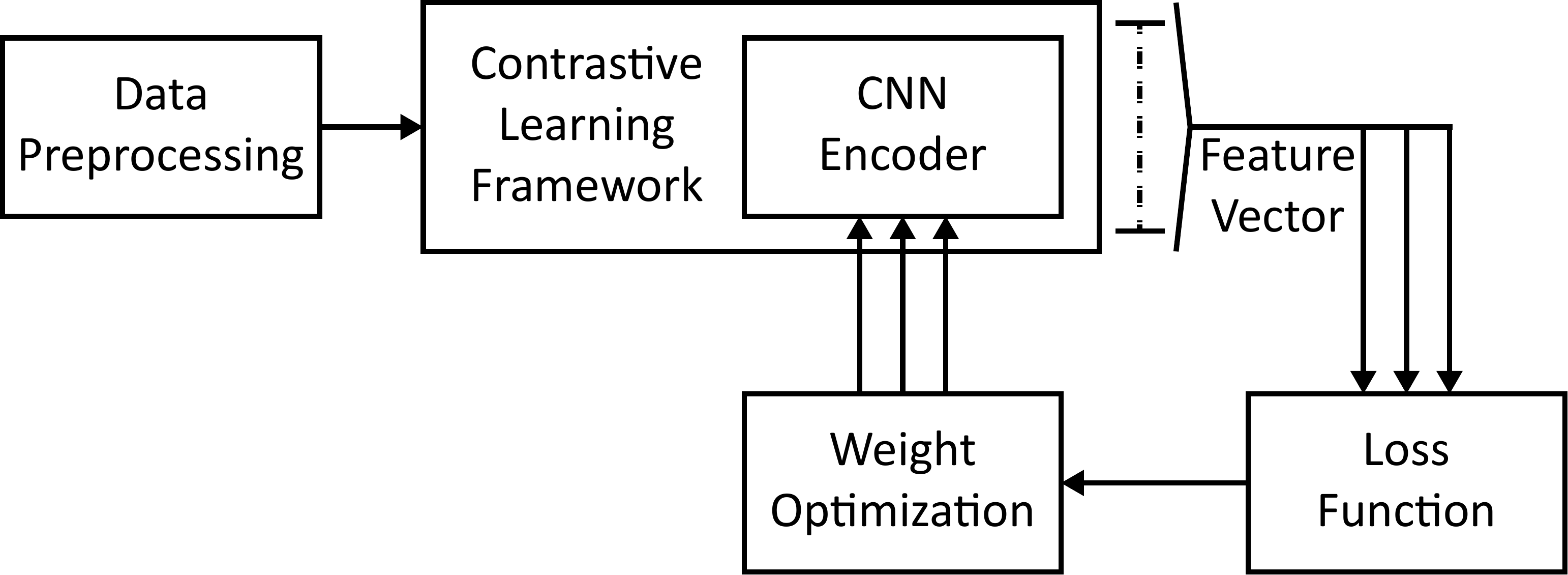}
    \caption{{Contrastive learning-based training.}}
    \label{2_Train}
\end{figure}

\begin{figure}[ht]
    \centering
    \includegraphics[width=\linewidth]{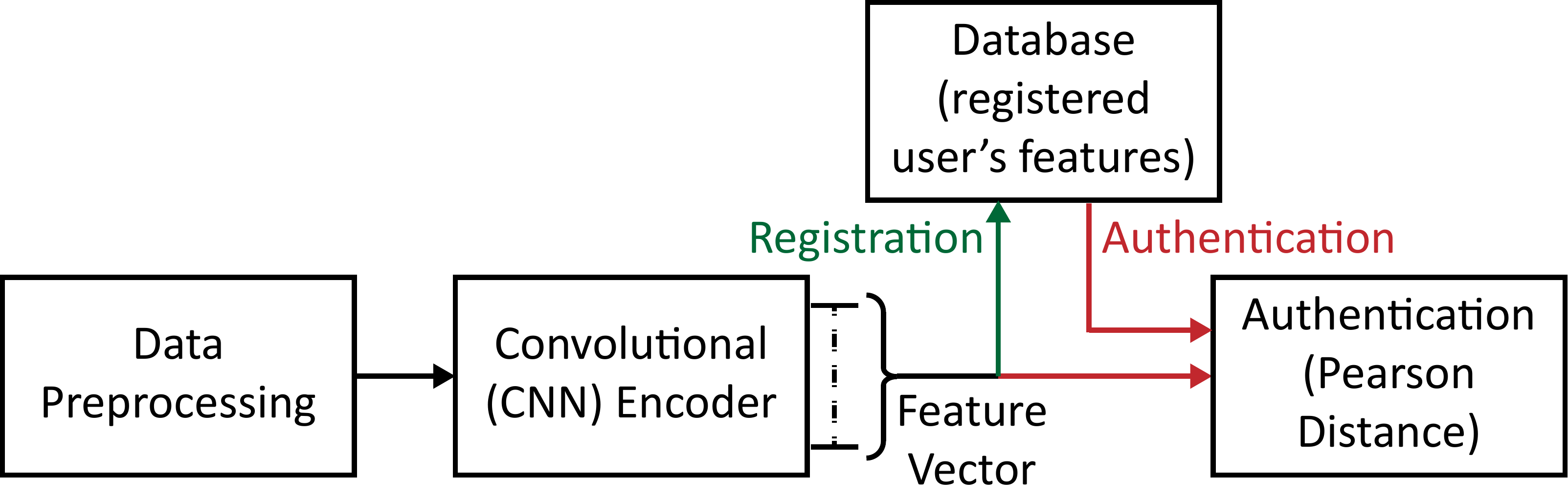}
    \caption{{Block diagram for user registration and authentication.}}
    \label{2_Auth}
\end{figure}

This section introduces the CNN-based user authentication system. The system involves a training step and an authentication step. \textit{Figure \ref{2_Train}} shows the workflow of training. Multiple original unlabelled ECG signals are the inputs of the data preprocessing module; the contrastive learning framework generates feature vectors by inferencing preprocessed signals. A loss is computed to update the CNN encoder's weight. The details are discussed in Section \ref{sec: training}. The authentication part consists of the following major modules: (1) data preprocessing, (2) feature extraction, (3) distance calculation, and (4) authentication. During registration, specified users are designated as authenticated users. The system generates a vector representing unique features for each authenticated user and will store it in a database. A corresponding feature vector is generated when a (new) user tries to log in. Further, the Pearson Correlation Coefficient between the generated vector and vectors of authenticated users stored in the database is computed to obtain a matching score. The system decides on the success of authentication by comparing the matching scores with a threshold. \textit{Figure \ref{2_Auth}} shows the authentication steps; details are discussed below.

\subsection{Data preprocessing}
This step aims to reduce noise and segment the ECG for further processing. Specific steps performed are as follows:

\subsubsection{Noise Reduction}
ECG acquired using non-invasive sensors often contains various noises such as baseline wander ($0.5$ to $0.6$ Hz), high-frequency muscle noise, and powerline noise. We used a bandpass filter with a low cutoff frequency of 0.5Hz and a high cutoff frequency of 40Hz to remove the baseline wander and muscle noise, respectively.

\subsubsection{Segmentation}
Regions of ECG that may not significantly contribute are removed to reduce the size of the input vector fed to the CNN feature extractor. 
RR and PT interval are generally used in feature extraction \cite{baranchuk2015p}, so we propose three different ECG segmentation methods and compare performances. 


\paragraph{R-peak to R-peak (R2R)}
This method detects all R peaks of the complete ECG signal with a Pan Tompkins QRS Detector\cite{pan1985real}, then gets the samples between the two R peaks as a piece. A value is settled to limit the piece length. Pieces over this piece length are ignored. Five consecutive pieces are spliced into one segment after resampling each piece to $200$ samples. Each piece consists of the previous beat's second half, the next beat's first half, and the adjacent beats' intervening regions. The combination of five consecutive pieces was done to reduce the impact of outliers introduced by Peak Detection \cite{labati2019deep}. {This method is simple to implement because of the distinctive R-peak characteristics of ECG. However, the information about the individual beats may be corrupted due to the independent resampling of each piece. Likewise, the potential errors in the QRS peak detection (accuracy of $99.3\%$ for R-peak detection) may result in losing the final authentication accuracy.}

\paragraph{P-peak to T-peak (P2T)}
This method detects P and T peaks of the ECG signal after QRS detection with a delineator \cite{martinez2004wavelet}, and the signal between these peaks of the same beat is considered a single segment. Based on the same idea in R2R, five such consecutive pieces are combined into one segment after resampling each piece to $200$ samples. {Compared with the R2R method, this method ensures that each segment in the instance is from a complete beat period. However, incorrect P- and T-peak detection introduce an additional error rate on top of the R-peak detection, resulting in more significant outliers. Therefore, the generated instances will be less than the R2R methods. Still, the potential loss of accuracy and increased detection complexity remains (accuracy of $99.6\%$ for P- and T-peak detection).}

\paragraph{No Peak Detection (NPD)}
This method randomly intercepts a fixed-length window ($1000$ samples) from the complete ECG signal after resampling, and each segment contains a variable number of ECG beats. We used the sampling rate of $200Hz$ and selected $1000$ samples to make the input form consistent with R2R. {As no detection and reassembly operations are performed, this method is the simplest to implement, with minimal complexity and no loss of information between adjacent beats. In addition, this method allows for the most significant number of original ECG samples to be retained because each segment is randomly intercepted.}

\begin{figure}[ht]
    \centering
    \includegraphics[width=0.7\linewidth]{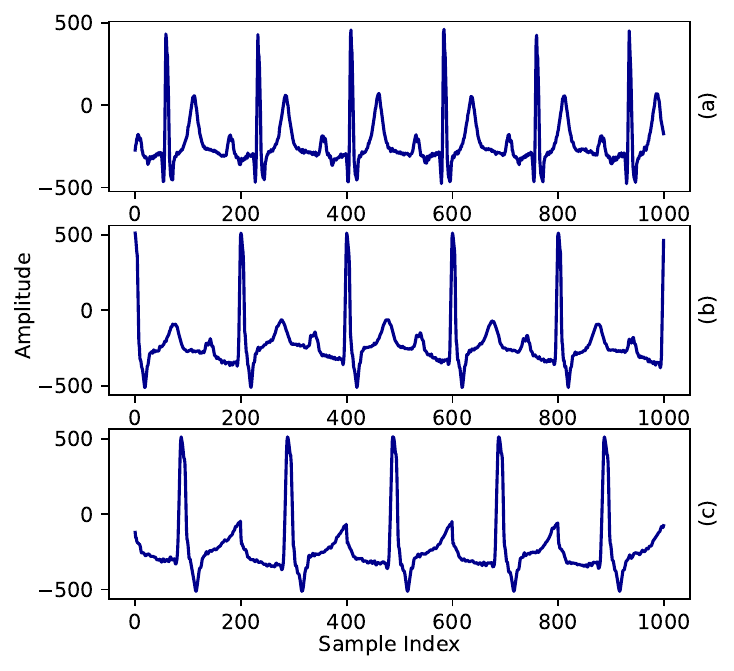}
    \caption{{Signal segmentation with different methods. (a) No Peak Detection, (b) R-peak to R-peak, (c) P-peak to T-peak.}}
    \label{4_segments}
\end{figure}

After resampling and normalization, the amplitude of segments lies in the range $[-512, 512]$. \textit{Figure \ref{4_segments}} shows the waveform of the ECG signal using the three different data segmentation methods, which is the output of the Data Processing module.

\subsection{Feature Extraction}
\begin{figure}[ht]
    \centering
    \includegraphics[width=\linewidth]{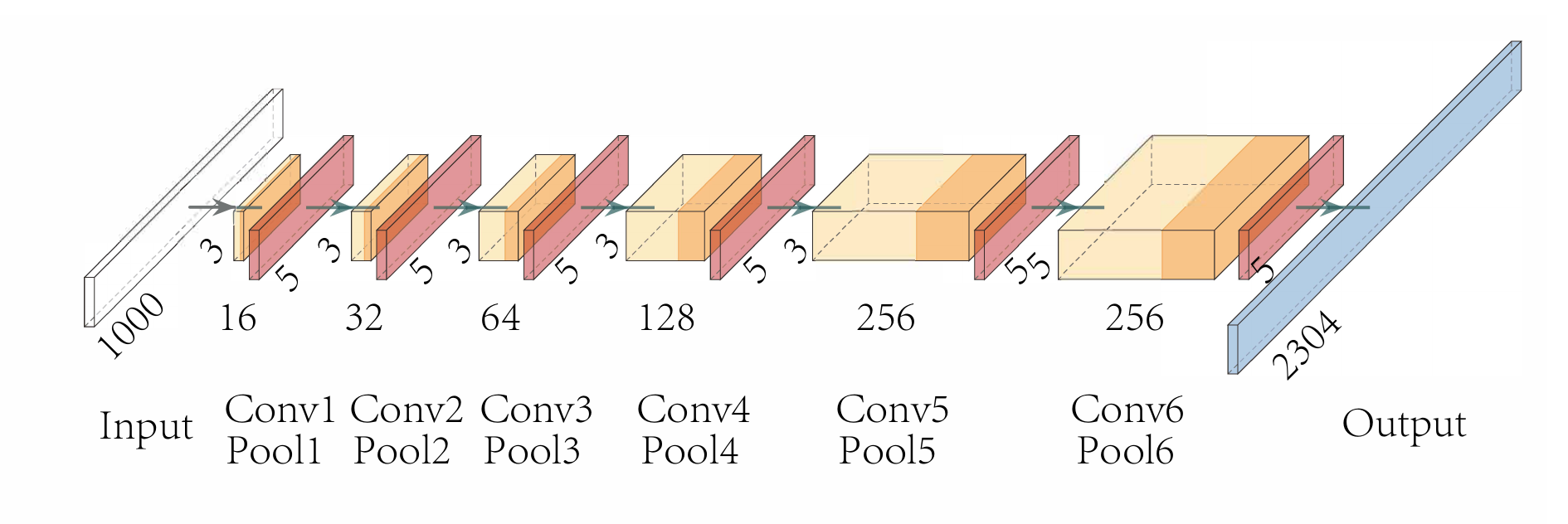}
    \caption{CNN Encoder used for feature extraction in the proposed Contrastive learning framework.}
    \label{5_cnn_stru}
\end{figure}

A CNN encoder trained from an ECG dataset extracts features from the segmented ECG signal. Since each processed ECG segment has a fixed number of samples in a segment ($1000$ samples) mimicking a time-domain snapshot, a 1D-CNN architecture is adopted to capture the features. The basic structure of the network is inspired from \cite{labati2019deep}, with fine-tuning applied to simplify the network structure and improve efficiency. The proposed architecture is shown in \textit{Figure \ref{5_cnn_stru}}. The architecture consists of six one-dimensional convolutional layers employing Rectified Linear Units (ReLU) neurons (yellow and orange blocks), enhancing non-linearity in the network and facilitating deeper representations. Additionally, six max-pooling layers (red blocks) are incorporated to down-sample the features, followed by a fully connected layer in the final layer. The last layer flattens the features into a vector, contributing to the comprehensive design of the network. The dimensions of each layer are detailed. After the network, a $1 \times 1000$ input will be extracted to a $1 \times 2034$ vector.

\subsection{Authentication}
The output of the last convolution layer is a $1 \times 2304$ vector, which contains ECG features that we use for authentication. During registration, ECG from authentic users are used to generate the inference, and the corresponding unique high-dimensional feature vectors generated by the CNN encoder are stored in a database. When the system receives a new authentication request, ECG from that user is used to generate a similar feature vector, which is then compared with those already stored in the database to compute a mapping score (Pearson Correlation Coefficient) as below (Eqn 1),

\begin{equation}
	D(X,Y) = \left|\frac{\sum_{i}(X_i-\bar{X})(Y_i-\bar{Y}) }{\sqrt{\sum_{i}(X_i-\bar{X})^2}\sqrt{\sum_{i}(Y_i-\bar{Y})^2}}\right|,
 \label{equ: PCC}
\end{equation}

Where $D(X, Y)$ is the mapping score, $X_i$-s are the feature values of the user under test, and $Y_i$-s are the feature values of a registered user within the database. A threshold value was determined during training. If $D(X, Y)$ is higher than the threshold, the user is authenticated and gets access to the system. Pearson Correlation Coefficient reflects the similarity of trends and shapes between samples. It is more suitable for signals than distances that depend on absolute values, such as the Euclidean and Hamming distance. Pearson Correlation Coefficient is scale-invariant, meaning it is not affected by the absolute values of the variables. This is in contrast to Euclidean and Hamming distance, which can be sensitive to the scale of the data. In addition, the Pearson Correlation Coefficient measures the linear relationship between two variables. It is sensitive to both the strength and direction of the linear association. This makes it suitable for capturing more complex relationships.

\section{Contrastive Learning Training}\label{sec: training}
In traditional ECG authentication systems, a supervised training approach is typically employed \cite{tantawi2015wavelet, hammad2018multimodal, thentu2021ecg, sepahvand2021novel}. Existing methods use labelled datasets, while more and larger datasets are unlabelled. We propose to train the system using contrastive learning to utilize larger datasets and help derive features. We utilize contrastive loss rather than classifier methods because there are a limited number of users during training; it is hard to add a new user as this requires the whole system to be retrained. As discussed below, two contrastive learning frameworks are proposed.

\subsection{Siamese Framework}
\begin{figure*}[ht]
      \centering
     \includegraphics[width=\linewidth]{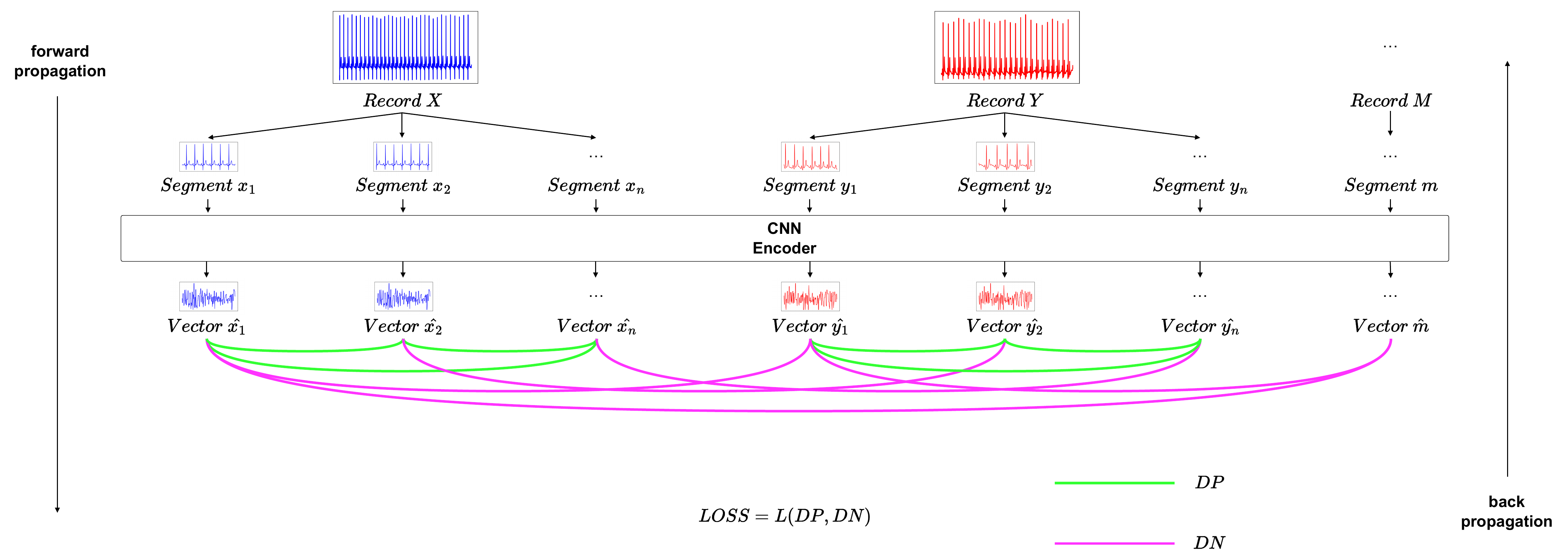}
     \caption{Illustration of siamese contrastive learning framework workflow in one epoch.}
     \label{7_siamese}
 \end{figure*}

In the siamese framework, two sets of segments are considered at once and compared with each other at every time step, and each set contains multiple segments from one record. The two sets could be similar (if taken from the same record, i.e. positive pair or dissimilar (if taken from different records, even if the same subject \cite{chen2020simple}, i.e. negative pair). \textit{Figure \ref{7_siamese}} shows the {flowchart} of the siamese framework. Every record will be started as a complete ECG in a batch with batch size $m\ (m \geq 2)$. Preprocessing, mentioned in Section III, is used to generate segments for different records. $n\ (n \geq 2)$ segments are selected randomly. This can be considered a kind of data augmentation, and the number of segments is predefined manually.

After data processing, the same CNN encoder transforms these generated segments into corresponding vectors. The vectors from the same records will be set as positive pairs, and vectors from different records will be set as negative pairs. Assuming that $n$ vectors are generated for each input, the number of positive pairs ($N_{pos}$) is:

\begin{equation}
\begin{aligned}
	N_{pos} &= C_n^2 \times m \\
	&= \frac{m \times n \times (n-1)}{2},
\end{aligned}
\end{equation}

where $C$ is the combination and $C_n^2=\frac{n\times (n-1)}{2}$. Moreover, the number of negative pairs ($N_{neg}$) is:

\begin{equation}
\begin{aligned}
	N_{neg} &= C_{m \times n}^2 - C_n^2 \times m \\
	&= \frac{m \times n \times (m \times n-1)}{2} - \frac{m \times n \times (n-1)}{2} \\
	&= \frac{m \times (m-1) \times n^2}{2}.
\end{aligned}
\end{equation}

As $m$ and $n$ increase, negative pairs will be much larger than positive pairs. $N_{neg}$ negative pairs will be randomly selected to reduce the training bias due to pair imbalance. This keeps the training from bias toward treating input pairs as positive or negative. Then, a loss function is applied:

\begin{equation}
	L = max\left(0,\;\overline{D_P}-\overline{D_N}+\lambda\right),
  \label{equ: sia}
\end{equation}

Where $\lambda$ is a manually set constant, the purpose of setting this constant here is that when the representation in a negative pair is good enough, as reflected by its distance, it is sufficiently far away. There is no need to waste time in that negative pair to increase that distance so that further training will focus on other more difficult-to-separate pairs. $\overline{D_P}$ and $\overline{D_N}$ are the average distances between positive and negative pairs. All the above distances are computed using Eqn \eqref{equ: PCC}. Because $L \geq 0$, $0 \leq \overline{D_P} \leq 1$ and $0 \leq \overline{D_N} \leq 1$, when we set $\lambda > 0.5$, this loss function will trend to make $\overline{D_P} < \lambda < \overline{D_N}$ and $\overline{D_N} - \overline{D_P} \geq \lambda$. Hence, the constant $\lambda$ set here will also be set as the threshold for the authentication stage.
\textit{Algorithm \ref{1_siamese_train}} summarizes the {pseudocode} of siamese contrastive learning technique.

\begin{algorithm}[ht]
\small
    \caption{Training Algorithm of Siamese Framework in One Epoch}
    \label{1_siamese_train}
        \KwData{training batch $B$, the number of segments for each record $N$, distance formula $D()$, constant $\lambda$}
        $M \leftarrow $ size of $B$\\
        \For{$m \leftarrow 0\ to\ M$}{
            \For{$n \leftarrow 0\ to\ N$}{
                $S_{mn} \leftarrow $ data processing on $B_m$
            }
            $X_{mn} \leftarrow $ model inferencing on $S_{mn}$\\
            $X_m \leftarrow \{X_{mn}\}$
        }
        \For{$i \leftarrow 0\ to\ M$}{
            \For{$j \leftarrow 0\ to\ M$}{
                $Mat_{ij} \leftarrow D(X_i, X_j)$
            }
        }
        $\overline{Mat_{pos}} \leftarrow $ average of $Mat_{pos}$\\
        $\overline{Mat_{neg}} \leftarrow $ average of $Mat_{neg}$\\
        $LOSS \leftarrow max(0,\overline{Mat_{pos}}-\overline{Mat_{neg}}+\lambda)$\\
        Encoder update
\end{algorithm}

\subsection{Triplet Framework}
In contrast to the siamese framework, the triplet framework also considers positive and negative pairs but generates them slightly differently.

\begin{figure*}[ht]
     \centering
     \includegraphics[width=\linewidth]{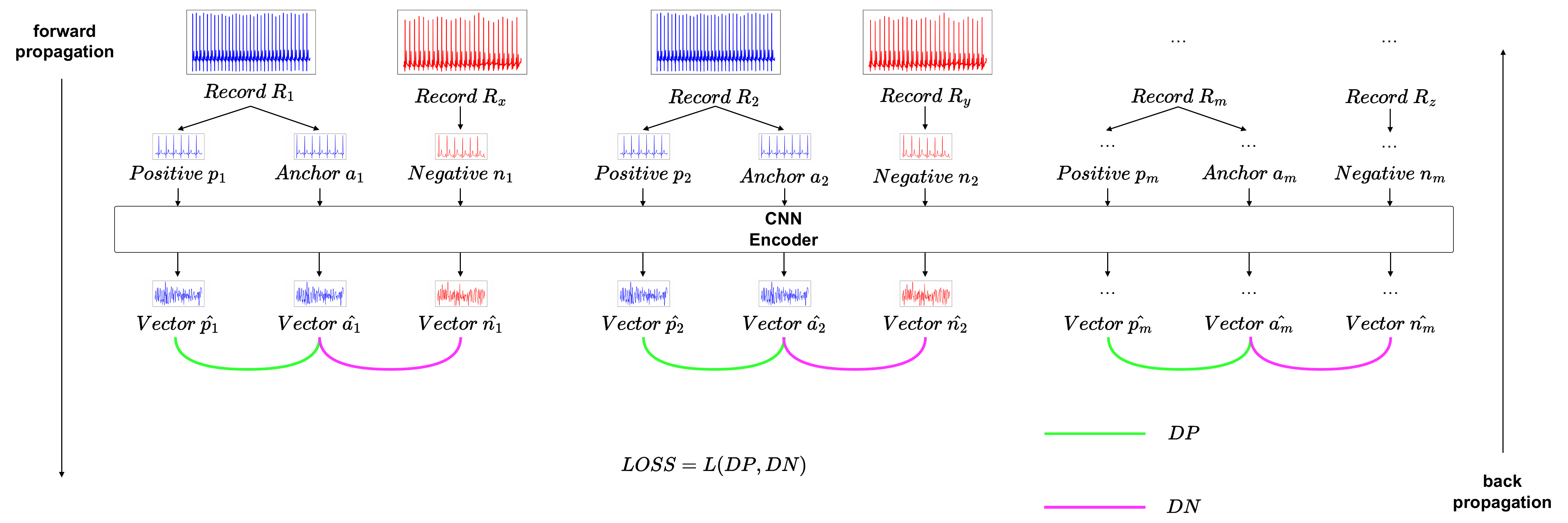}
     \caption{Illustration of triplet contrastive learning framework workflow in one epoch.}
     \label{8_triplet}
 \end{figure*}

\textit{Figure \ref{8_triplet}} shows the {flowchart} of the triplet framework. In a mini-batch with batch size $m\ (m \geq 2)$, every record will provide a complete ECG signal and extra $m$ ECG signals are selected randomly, which form $2m$ inputs in total. The $m$ inputs from the iterations for each record in this batch are called positive records, while the other $m$ inputs from the randomly selected records are called negative records.

Two segments are generated by the preprocessing method mentioned in Section III for each positive record. These two segments are set as anchor samples and positive samples. For each negative record, one segment is generated and called a negative sample. After feature extraction, a common CNN Encoder transforms these generated segments into the corresponding vectors. The vector from the positive sample and vector from the anchor sample form a positive pair, while the vector from the negative sample and vector from the anchor sample form a negative pair. One combination of these three vectors is called a triplet. The number of positive and negative pairs are equal:

\begin{equation}
	N_{pos} = N_{neg} = m.
\end{equation}

Then, a loss function is applied:

\begin{equation}
	L = \sum_{i = 1}^{m}{max\left(0,\;{D_P}_i-{D_N}_i+\lambda\right)},
 \label{equ: tri}
\end{equation}

Where $\lambda$ is a manually set constant and will also be set as the threshold for the authentication stage. ${{D_P}_i}$ and ${{D_N}_i}$ are the distance of the positive pair and negative pair for the $i$th triplet. 
Inverse Equation \eqref{equ: PCC} computes all the distances mentioned above, which makes a large value for the negative pair and a small value for the positive pair.

In the training process, there are three possible conditions of loss for a triplet \cite{pinto2020secure}:

\begin{itemize}
    \item easy triplets: ${D_N}_i > {D_P}_i + \lambda$. In this case, the negative and the anchor samples are already far apart in the embedding space compared to the positive sample. The loss is $0$, and the network parameters will not continue to update.
    \item hard triplets: ${D_N}_i < {D_P}_i$. In this case, the negative sample is closer to the anchor sample than the positive sample, and the loss is positive, so the network can continue to update.
    \item Semi-hard triplets: ${D_P}_i < {D_N}_i < {D_P}_i + \lambda$. The distance from the negative sample to the anchor sample is greater than the positive sample but does not exceed the set constant $\lambda$. The loss is still positive, and the network can continue to update.
\end{itemize}

Easy triplets should be avoided to be sampled as much as possible since their loss of $0$ does not help optimize the network. Since negative samples are randomly generated from negative records, this framework cannot perform negative sample selection, which can be solved by adjusting the batch size.

\textit{Algorithm \ref{2_triplet_train}} summarizes the {pseudocode} of triplet framework.

\begin{algorithm}[ht]
\small
    \caption{Training Algorithm of Triplet Framework in One Epoch}
    \label{2_triplet_train}
        \KwData{training batch $B$, distance formula $D()$, constant $\lambda$}
        $M \leftarrow $ size of $B$\\
        \For{$m \leftarrow 0\ to\ M$}{
            $i \leftarrow $ Randomly select a number from $0$ to $M$ but $ \neq m$\\
            ${S_A}_m \leftarrow $ data processing on $B_m$\\
            ${S_P}_m \leftarrow $ data processing on $B_m$\\
            ${S_N}_m \leftarrow $ data processing on $B_i$\\
            ${X_A}_m \leftarrow $ model inferencing on ${S_N}_m$\\
            ${X_P}_m \leftarrow $ model inferencing on ${S_N}_m$\\
            ${X_N}_m \leftarrow $ model inferencing on ${S_N}_m$
        }
        $D_P \leftarrow D({X_A}_m, {X_P}_m)$\\
        $D_N \leftarrow D({X_A}_m, {X_N}_m)$\\
        $L \leftarrow max(0,\sum{D_P}-\sum{D_N}+\lambda)$\\
        Encoder update
\end{algorithm}

\section{System Optimization}\label{sec: optim}
The proposed system will be deployed on a wearable edge sensor with limited performance and resources. In wearable sensors, multiple circuitry for data acquisition, processing, and transmission already exist, and therefore, the proposed technique, when integrated, must not significantly increase the overall power consumption. Consequently, we employed model compression techniques such as quantization and pruning to reduce the power and resource utilization of the model for hardware constraint adaptation, faster inferencing, and longer battery life, as described below.

\subsection{Model Quantization}
The convolutional operations in CNNs require several floating-point (FP) arithmetic operations.
The implementation complexity and power consumed by FP arithmetic are higher in custom hardware or embedded MCUs. In addition, not all MCUs have an integrated FP unit. To address this, we experimented by replacing the FP weights in the original model with quantized values at different quantization levels \cite{gholami2022survey}. 

High-bit-width floating-point weights are converted into low-bit-width fixed-point weights using the following function:

\begin{equation}
	X \approx \frac{round(2^n \cdot X)}{2^n} = \frac{Y}{2^n},
\end{equation}

where $X$ is a high-bit-width floating-point number (original weight), and $Y$ is the low-bit-width quantized weight and an integer. $n$ is a hyperparameter that controls the precision of quantization. As $n$ increases, the quantized weights become closer to the original weights.

Also, quantization accelerates neural network inference by converting the calculation method. It can be proven that any integer can be represented as the sum of powers-of-two ($Y=2^a+2^b+\dots$). Hence, multiplications can be transformed into additions and bit-shift operations \cite{wang2020low}. Thus, $N \times X$ can be simplified to:

\begin{equation}
\begin{aligned}
	N \times X &\Rightarrow N \times \frac{Y}{2^n}\\
	\Rightarrow N \times \frac{2^a+2^b+\dots}{2^n} &\Rightarrow N\cdot2^{a-n}+N \cdot 2^{b-n}+\dots,
\end{aligned}
\end{equation}

where $N\cdot2^{a-n}$ means bit-shift operations for $N$ if $a$ and $n$ are integers and the equation follows binary arithmetic. $a-n > 0$ corresponds $N$ left-shifts $a-n$ bits and $a-n < 0$ corresponds $N$ right-shifts $a-n$ bits. For example, if $N=2$ ($10$ in binary) and $a-n=3$, the result of $N\cdot2^{a-n}$ should be $2=10(b)\ll3=10000(b)=16$.

\subsection{Model Pruning}
Quantization optimizes the neural network in terms of data types. Besides, the neural network structure is worth optimization. As mentioned in Section \ref{sec: optim}-A, there are many redundant parameters in the neural network, and the model can be pruned by deleting these parameters, thereby reducing the model size and time complexity \cite{fernandes2021pruning}.

Pruning optimization is performed during training instead of after training to minimize the loss of precision caused by pruning \cite{roy2020pruning}. We tested different sparsity values to achieve a balance between accuracy and complexity.

\section{Evaluation and Results}\label{sec: results}
This section presents the datasets and platforms used in this work. In addition, the performance achieved under different conditions is discussed.

\subsection{Datasets}
We used three different datasets in this work. One dataset was used to train the model and evaluate testing, and the remaining two were used exclusively to test the model's performance. PTB Diagnostic ECG Database (PTBDB) \cite{bousseljot1995nutzung}, which is the most popular database used for ECG-based authentication tasks, was used for training and testing the model \cite{tantawi2015wavelet, hammad2018multimodal, thentu2021ecg, prakash2022baed, sepahvand2021novel, hazratifard2023ensemble}. This database contains $549$ records from $290$ subjects. 12 leads of each record were used.
The signal is digitized at $1000$Hz, with a $16$-bit resolution over a $\pm 16.384$ mV range. All records were used for training, and specified authenticated subjects had their feature vector extracted and stored for the authentication step. 
Many machine learning models reported in the literature are often trained and tested using the same database \cite{tantawi2015wavelet, hammad2018multimodal, thentu2021ecg, sepahvand2021novel}. They are inherently brittle when tested with different datasets due to the overfitting of the model to various signal parameters. To counter this issue and ensure our model is generalizable, we tested our model with datasets that were \textbf{\textit{previously unseen}} by the model during the training process. We used the previously unseen MIT-BIH Arrhythmia Database (MITDB) \cite{moody2001impact} to test the model's performance. This database contains $48$ two-channel ambulatory ECG records from $47$ subjects studied by the BIH Arrhythmia Laboratory between 1975 and 1979, and all subjects were used for testing.
The signals are digitized at $360$ Hz, with $11$-bit resolution over a $\pm 5$ mV range. MITDB is commonly used in ECG-related research and has persons' identity labels. We also used another unseen ECG-ID database (ECGIDDB) \cite{lugovaya2005biometric} to evaluate the model performance. This database contains $310$ records from $90$ subjects. Each record includes raw and filtered signals, and all subjects were used for testing.
The signals are digitized at $500$ Hz, with $12$-bit resolution over a $\pm 10$ mV range. ECGID is collected explicitly for identification and authentication tasks, making the experiment's implementation easier. The databases are widely used in ECG-related research and collected from natural persons. {In addition, these datasets come with varying degrees of noise, which can be used to test the system's immunity to interference.} We selected the same datasets used in recent work and conducted more experiments in multiple frameworks and different settings. In addition, each subject generated segments at the beat level, which contains different situations (multiple traces, different periods, normal and abnormal beats) and ensures the universality of databases.

\subsection{Experiment Setup}
The baseline system is implemented on PyTorch with a PC with $10$ cores Intel (R) Core (TM) i9-10900f CPU, RAM $64$ GB, GPU NVIDIA RTX3090 with $24$ GB of memory. Stochastic Gradient Descent (SGD) trains the CNN Encoder with a batch size of $256$ for $1000$ epochs. Adam optimizer was used. The initial learning rate is $0.05$ and decreased with Cosine Annealing. We set the training parameter $\lambda = 0.7$ for both frameworks.

\begin{figure}[ht]
    \centering
    \includegraphics[width=\linewidth]{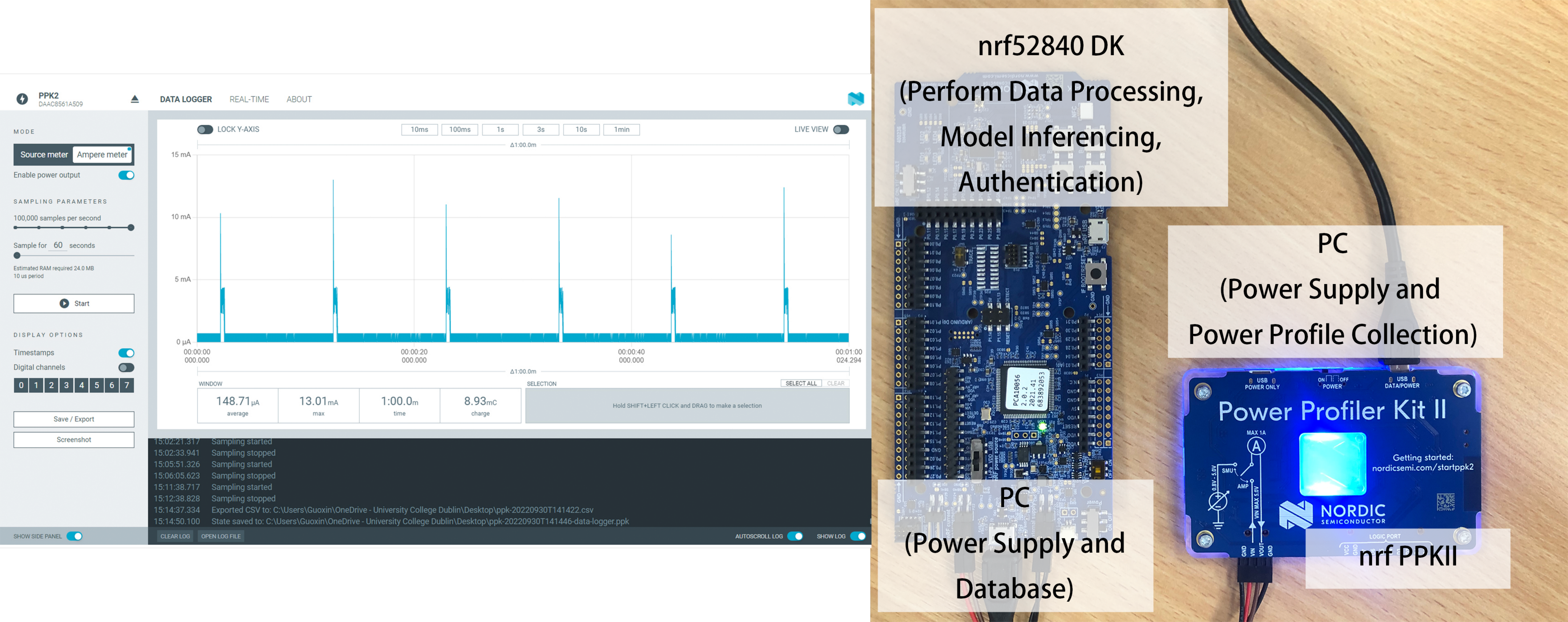}
    \caption{Experimental setup for power measurements}
    \label{11_platform}
\end{figure}

After encoder training, the model is optimized as discussed in section V. INT4, INT8 and INT16 data formats were set in quantization optimisation to test the effect of different parameters on the optimised performance. Similarly, sparsity of $20\%$, $40\%$ and $80\%$ were used for pruning.
We used nRF52840DK, which integrates nRF52 Bluetooth LE SoC from Nordic Semiconductor as the experimental platform for deploying the model. nRF52 SoC is built around the $32$-bit ARM Cortex-M4 CPU with a floating-point unit running at $64$ MHz. It has $1$ MB of Flash and $256$ KB of RAM. The CMSIS-NN library implemented the neural network deployment.
When deployed on nRF52840DK, the model power consumption is measured using nRFPPKII (Power Profiler Kit II). The embedded experimental environment is shown in \textit{Figure \ref{11_platform}}.

In the experiments on the embedded platform, we start by deploying the optimized models as mentioned in Section \ref{sec: optim}. During the registration process, several subjects are selected as authentic users. The ECGs of these users are sent to the nRF52 board via UART, and after model inference, the feature vectors are saved and recorded as a database in the SRAM memory of nRF52. During the authentication process, data from test subjects are fed into the same preprocessing and inferencing logic. Finally, the extracted features are compared with the pre-stored values in SRAM, and the authentication result of the subject is returned. We measured the power consumption during the authentication loop and computed the average and peak power.

\begin{figure*}[ht]
    \centering
    \begin{subfigure}[b]{0.26\linewidth}
        \centering
        \includegraphics[width=\linewidth]{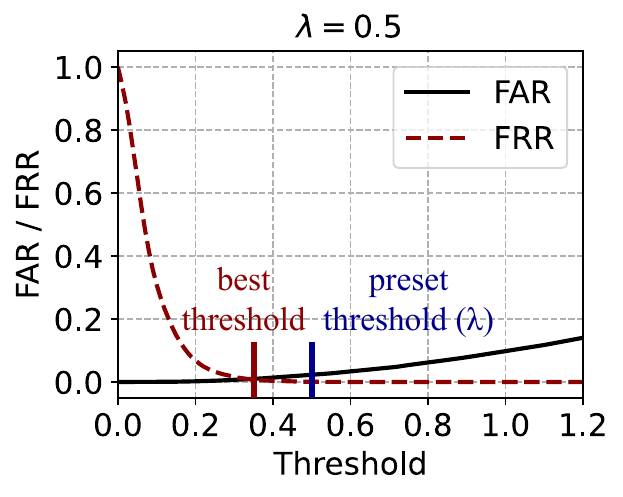}
        \caption{}
    \end{subfigure}
    \begin{subfigure}[b]{0.26\linewidth}
        \centering
        \includegraphics[width=\linewidth]{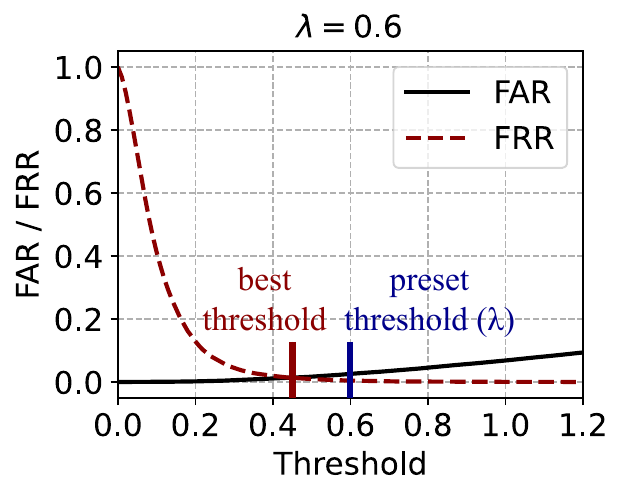}
        \caption{}
    \end{subfigure}
    \begin{subfigure}[b]{0.26\linewidth}
        \centering
        \includegraphics[width=\linewidth]{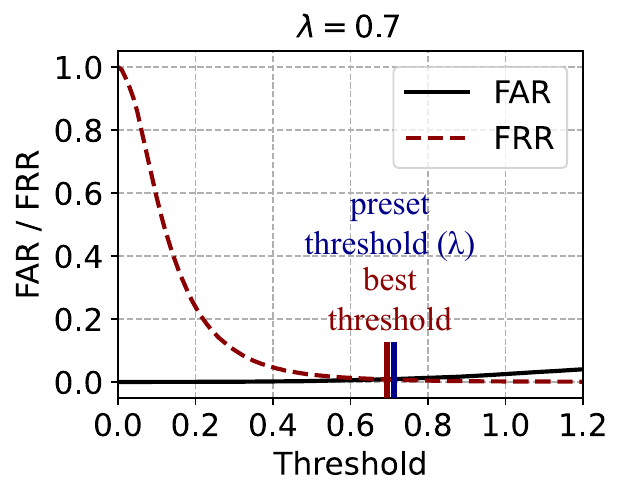}
        \caption{}
    \end{subfigure}\\
    \begin{subfigure}[b]{0.26\linewidth}
        \centering
        \includegraphics[width=\linewidth]{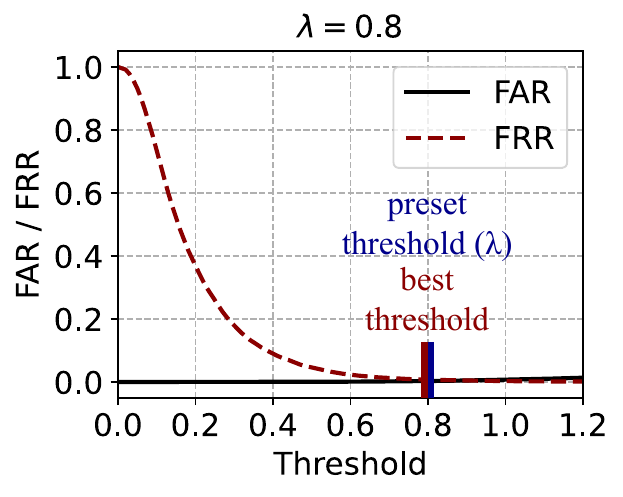}
        \caption{}
    \end{subfigure}
    \begin{subfigure}[b]{0.26\linewidth}
        \centering
        \includegraphics[width=\linewidth]{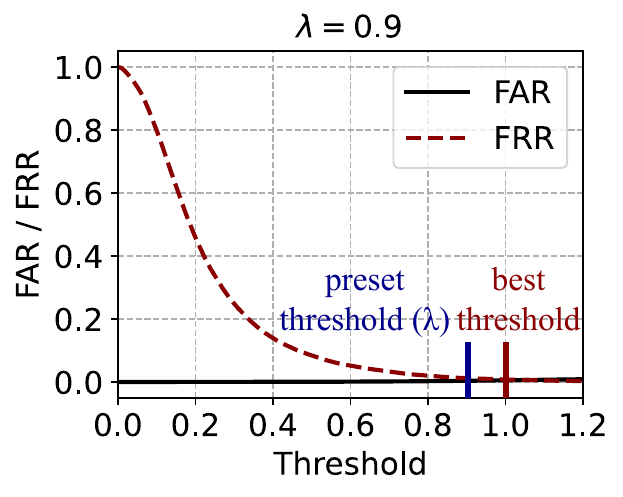}
        \caption{}
    \end{subfigure}
    \begin{subfigure}[b]{0.26\linewidth}
        \centering
        \includegraphics[width=\linewidth]{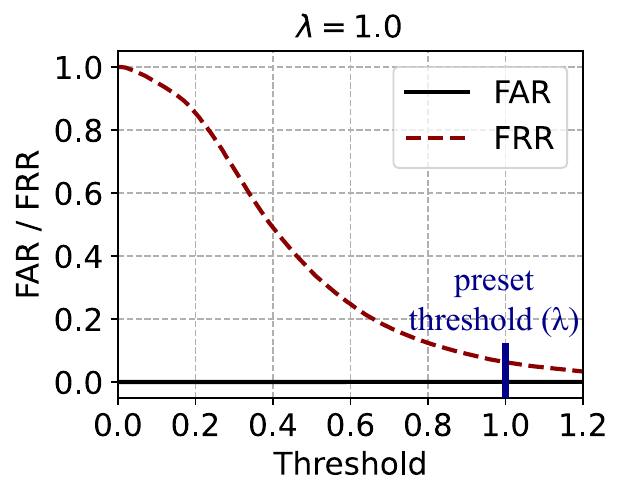}
        \caption{}
    \end{subfigure}
    \caption{FAR, FRR plots for varying $\lambda$. (a) $\lambda=0.5$. (b) $\lambda=0.6$. (c) $\lambda=0.7$. (d) $\lambda=0.8$. (e) $\lambda=0.9$. (f) $\lambda=1.0$.}
    \label{8_FAR_FRR}
\end{figure*}

\subsection{Metrics}
\subsubsection{System Performance}
System performance is the most critical metric for the authentication system. System performance is broken down into various considerations: Precision, Recall, False Acceptance Rate (FAR), False Rejection Rate (FRR), True Positive Rate (TPR), False Positive Rate (FPR), F1 score and Accuracy \cite{hossin2015review}. We set one subject as a registered user and the remaining as unregistered users. Then, we randomly generated 100 segments from the registered users' records and 100 segments from unregistered users' records for testing. After testing each subject once in the dataset, we counted true positives (TP), true negatives (TN), false positives (FP), and false negatives (FN). We performed the above evaluation once for each subject in the dataset and summarized the results obtained. Then, we calculated the other metrics based on them.

\subsubsection{Time Complexity}
For an embedded system, response time is usually represented by the number of operations. According to different CPU architectures, the number of operations can be replaced by the CPU cycle.

\subsubsection{Average Power}
For an embedded system, energy consumption is closely related to lifetime, and the average power value at constant voltage input is used to reflect the power consumption.

\subsection{Evaluation}
\subsubsection{Accuracy on PTBDB}
\textit{Table \ref{1_processing_framework}} presents the accuracy of using different preprocessing methods and frameworks on the PTBDB dataset.

\begin{table}[ht]
\centering
\captionsetup{justification=centering}
\caption{Comparison between different preprocessing methods and training frameworks}
\scriptsize
\label{1_processing_framework}
\resizebox{\columnwidth}{!}{%
\begin{tabular}{ccccc}
\hline
\begin{tabular}[c]{@{}c@{}}Segmentation\\ Method\end{tabular} & \begin{tabular}[c]{@{}c@{}}Training\\ Framework\end{tabular} & \begin{tabular}[c]{@{}c@{}}Prec\\ (\%)\end{tabular} & \begin{tabular}[c]{@{}c@{}}Recall\\ (\%)\end{tabular} & \begin{tabular}[c]{@{}c@{}}Accuracy\\ (\%)\end{tabular} \\ \hline
NPD & Siamese & 96.33 & 98.70 & 97.53 \\
NPD & Triplet & 98.52 & 99.78 & 99.15 \\
R2R & Siamese & 95.81 & 96.37 & 96.10 \\
R2R & Triplet & 97.06 & 98.04 & 97.56 \\
P2T & Siamese & 98.04 & 97.67 & 97.85 \\
P2T & Triplet & 99.28 & 99.04 & 99.16 \\ \hline
\end{tabular}%
}
\end{table}

The evaluation randomly divides the dataset into unregistered users and registered users for login tests. The matching score is recorded in each login operation. Finally, total accuracy is obtained.

The results show that the triplet framework outperforms the siamese framework in all metrics, and the P2T method provides the highest accuracy in preprocessing. 
In addition, the number of heartbeats in the segments generated by the P2T method is constant and aligned, making the final determination more accurate. As for the R2R method, although the number of segmented heartbeats it generates is also constant, this approach combines the second half of the last heartbeat and the first half of the current heartbeat, which could destroy the completion of a single ECG beat and have resulted in lower performance. The NPD method performs well even though segmentation is random and results in distinct patterns because the neural network learned extra information, e.g. frequency and amplitude. The resulting segment length is large enough to contain several complete heartbeats. 

\subsubsection{Accuracy on MITDB and ECGIDDB}
We used previously unseen MITDB and ECGIDDB to evaluate the proposed system's generalization ability. These tests uniformly use the triplet framework but with different preprocessing methods. The encoders are only trained using the PTBDB and are directly used for testing on other databases without changing any parameters, thresholds, and network structures. \textit{Table \ref{2_processing_dataset}} presents the accuracy of using different preprocessing methods and datasets.

\begin{table}[ht]
\centering
\captionsetup{justification=centering}
\caption{Comparison between different preprocessing methods and testing datasets}
\label{2_processing_dataset}
\resizebox{\columnwidth}{!}{%
\begin{tabular}{ccccc}
\hline
\begin{tabular}[c]{@{}c@{}}Segmentation\\ Method\end{tabular} & \begin{tabular}[c]{@{}c@{}}Test\\ Set\end{tabular} & \begin{tabular}[c]{@{}c@{}}Prec\\ (\%)\end{tabular} & \begin{tabular}[c]{@{}c@{}}Recall\\ (\%)\end{tabular} & \begin{tabular}[c]{@{}c@{}}Accuracy\\ (\%)\end{tabular} \\ \hline
NPD & MITDB & 98.24 & 99.09 & 98.67 \\
NPD & ECGIDDB & 98.32 & 99.21 & 98.77 \\
R2R & MITDB & 92.45 & 94.34 & 93.45 \\
R2R & ECGIDDB & 92.90 & 95.03 & 94.02 \\
P2T & MITDB & 94.06 & 95.28 & 94.70 \\
P2T & ECGIDDB & 95.30 & 97.68 & 96.52 \\ \hline
\end{tabular}%
}
\end{table}

Due to the differences in data composition, acquisition method, and acquisition platform of each dataset, which affects the system performance to different degrees, the NPD method achieves the highest accuracy and the best generalization ability in this experiment; this method does not rely on peak detection, which reduces the processing complexity further.




\subsubsection{Threshold Selection}
\textit{Figure \ref{8_FAR_FRR}} presents the graph between the authentication threshold and false acceptance rate (FAR) / false rejection rate (FRR) of the results obtained with different training constant $\lambda$ values using the NPD and PTBDB on the triplet framework. The threshold value significantly influences FAR and FRR and, therefore, is chosen to minimise these values. As mentioned in IV, the threshold is determined at the training phase, and we used the same value of constant $\lambda$ as the threshold. {The preset threshold should be close to the best threshold. Therefore we recommend setting $\lambda$ between $0.7$ and $0.8$.}

\subsubsection{Receiver Operator Characteristic Curve}
\begin{figure}[ht]
    \centering
    \begin{subfigure}[b]{0.45\linewidth}
        \centering
        \includegraphics[width=\linewidth]{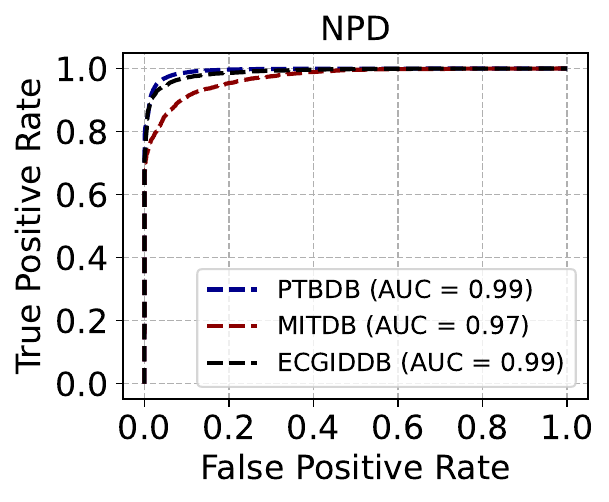}
        \caption{}
        \label{12_npd_roc}
    \end{subfigure}\\
    \begin{subfigure}[b]{0.45\linewidth}
        \centering
        \includegraphics[width=\linewidth]{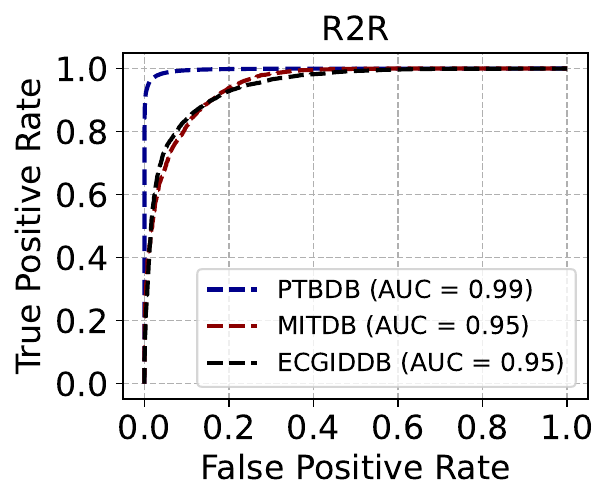}
        \caption{}
        \label{13_r2r_roc}
    \end{subfigure}
    \begin{subfigure}[b]{0.45\linewidth}
        \centering
        \includegraphics[width=\linewidth]{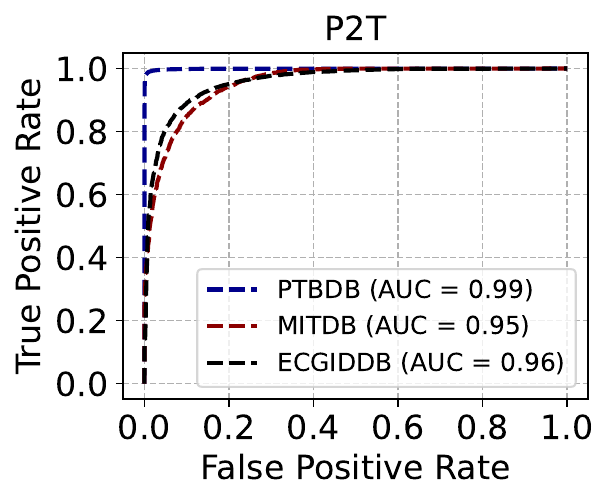}
        \caption{}
        \label{14_p2t_roc}
    \end{subfigure}
    \caption{ROC curve of different preprocessing methods on 3 datasets. (a) NPD method. (b) R2R method. (c) P2T method.}
    \label{9_ROC}
\end{figure}

We used the Receiver Operator Characteristic (ROC) curve to present the authentication system's generalization performance; its vertical axis is the True Positive Rate (TPR), and its horizontal axis is the False Positive Rate (FPR). \textit{Figure \ref{9_ROC}} presents the ROC curve of different preprocessing methods on different datasets with the triplet framework.

The Area Under Curve (AUC) is the area under the ROC curve, and the larger the AUC, the better the performance. This experiment shows that all three preprocessing methods achieved good results on PTBDB ($0.99$). The NPD method achieved the AUC of $0.97$ on MITDB and $0.99$ on ECGIDDB, which shows good generalizability.

\subsubsection{Repeat Authentication}
Realistic authentication systems, such as mobile devices, usually prevent unregistered users from getting through while allowing several attempts by the registered user. We experimented with authenticating the user multiple times repeatedly so that the user would be granted access if authentication was successful at least once. {A user will record ECG multiple times to get different segments with the same processing method.} This increased the accuracy for all datasets, reaching nearly $100\%$ in the PTBDB and ECGIDDB with three repeat attempts. \textit{Table \ref{3_repeat}} presents the experimental results using different preprocessing methods and datasets with the triplet framework in this condition.

\begin{table}[ht]
\scriptsize
\centering
\captionsetup{justification=centering}
\caption{Comparison between different preprocessing methods and testing datasets with different repeat times}
\label{3_repeat}
\resizebox{\columnwidth}{!}{%
\begin{tabular}{ccccc}
\hline
\begin{tabular}[c]{@{}c@{}}Segmentation\\ Method\end{tabular} &
  \begin{tabular}[c]{@{}c@{}}Repeat\\ Times\end{tabular} &
  \begin{tabular}[c]{@{}c@{}}PTBDB Acc\\ (\%)\end{tabular} &
  \begin{tabular}[c]{@{}c@{}}MITDB Acc\\ (\%)\end{tabular} &
  \begin{tabular}[c]{@{}c@{}}ECGIDDB Acc\\ (\%)\end{tabular} \\ \hline
\multirow{3}{*}{NPD} & One   & 99.15 & 98.67 & 98.77 \\
                     & Two   & 99.90 & 99.19 & 99.21 \\
                     & Three & 100   & 99.80 & 99.83 \\
\multirow{3}{*}{R2R} & One   & 97.56 & 93.45 & 94.02 \\
                     & Two   & 99.76 & 95.40 & 97.45 \\
                     & Three & 99.93 & 97.52 & 99.30 \\
\multirow{3}{*}{P2T} & One   & 99.16 & 94.70 & 96.52 \\
                     & Two   & 99.97 & 95.55 & 98.34 \\
                     & Three & 100   & 97.45 & 99.62 \\ \hline
\end{tabular}%
}
\end{table}

\subsubsection{Optimization Results}
\begin{figure}[ht]
    \centering
    \includegraphics[width=\linewidth]{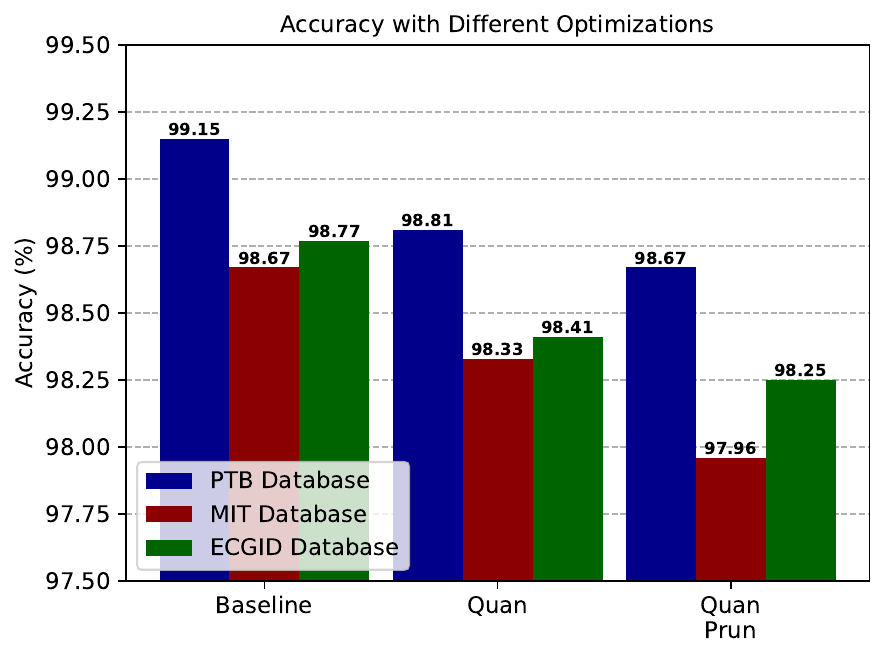}
    \caption{Accuracy after optimization for three datasets.}
    \label{15_optim_acc}
\end{figure}

\textit{Figure \ref{15_optim_acc}} presents the accuracy of different optimization methods and datasets. The results show that the baseline system has the highest accuracy ($99.15\%$) because it uses original weights with higher precision that contain full information. 
The quantized system achieves a lower accuracy ($98.81\%$).
The degree of quantization is set to $8$ ($n=8$) to compress the model size and increase the inference speed while maintaining accuracy. The pruned system achieves lower accuracy than the quantized system because insignificant connections and neurons are eliminated in the pruning process, but the accuracy remains high ($98.67\%$). In this experiment, sparsity was set to $20\%$. The test on MITDB achieves a similar result, which shows the system's strong generalization ability. \textit{Table \ref{4_optim_acc}} shows the precision, recall, F1-score and accuracy with different optimizations on PTBDB.

\begin{table}[ht]
\centering
\captionsetup{justification=centering}
\caption{Accuracy for different optimizations on PTBDB}
\label{4_optim_acc}
\resizebox{\columnwidth}{!}{%
\begin{tabular}{cccc}
\hline
Metric &
  \begin{tabular}[c]{@{}c@{}}Baseline Syetem\\ (\%)\end{tabular} &
  \begin{tabular}[c]{@{}c@{}}Quantization\\ (\%)\end{tabular} &
  \begin{tabular}[c]{@{}c@{}}Quantization\& Pruning\\ (\%)\end{tabular} \\ \hline
Precision & 98.52 & 98.37 & 98.51 \\
Recall    & 99.78 & 99.27 & 98.83 \\
F1-score  & 99.15 & 98.82 & 98.67 \\
Accuracy  & 99.15 & 98.81 & 98.67 \\ \hline
\end{tabular}%
}
\end{table}

\begin{figure*}
    \centering
    \begin{subfigure}[b]{0.31\textwidth}
        \centering
        \includegraphics[width=\linewidth]{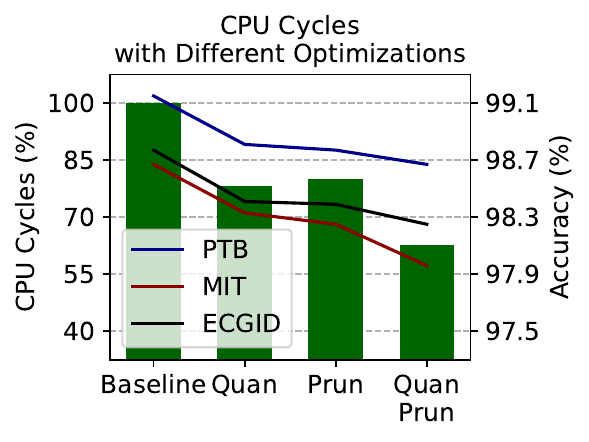}
        \caption{}
        \label{16_optim_time}
    \end{subfigure}
    \begin{subfigure}[b]{0.31\textwidth}
        \centering
        \includegraphics[width=\linewidth]{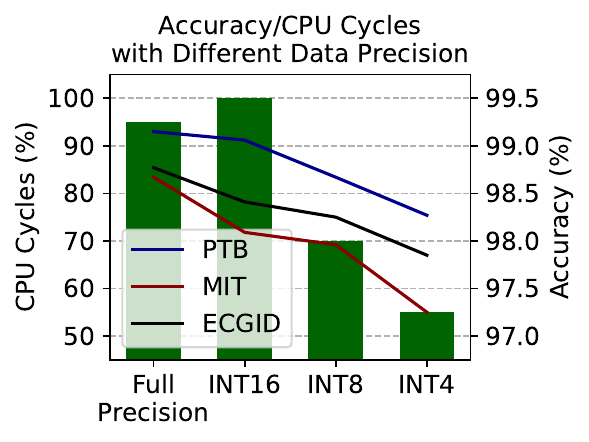}
        \caption{}
        \label{17_quan_cont}
    \end{subfigure}
    \begin{subfigure}[b]{0.31\textwidth}
        \centering
        \includegraphics[width=\linewidth]{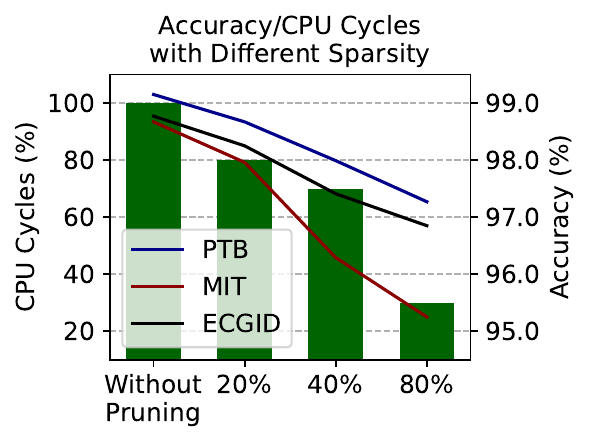}
        \caption{}
        \label{18_prun_cont}
    \end{subfigure}
    \caption{(a) The results of CPU cycles with different optimizations. (b) The results of accuracy and CPU cycles in different quantization coefficients. (c) The results of accuracy and CPU cycles in different pruning coefficients.}
\end{figure*}

\textit{Table \ref{5_optim_time}} presents the number of operations of the baseline system using various optimization methods, which indicates time complexity. The results show that the baseline system requires $459040$ multiplications, resulting in significant time complexity. The quantized and pruned system performs better because it uses bits-shift and additions instead of multiplications. Also, it removed unimportant nodes, which decreased the running time.

\begin{table}[ht]
\centering
\captionsetup{justification=centering}
\caption{Number of Operations after optimization}
\label{5_optim_time}
\resizebox{\columnwidth}{!}{%
\begin{tabular}{ccccc}
\hline
\multirow{2}{*}{System} & \multicolumn{4}{c}{Number of Operations}           \\ \cline{2-5} 
                        & Multiplication & Inversion & Bits-shift & Addition \\ \hline
Baseline                & 459040         & 0         & 0          & 460553   \\
Quantization            & 0              & 59112     & 459040     & 919593   \\
Pruning                 & 367232         & 0         & 0          & 368442   \\
Quantization \& Pruning & 0              & 47289     & 367232     & 735674   \\ \hline
\end{tabular}%
}
\end{table}

\begin{table*}[ht]
\scriptsize
\centering
\captionsetup{justification=centering}
\caption{Comparison of the proposed system with other published works}
\label{system_comparision}
\resizebox{0.9\textwidth}{!}{%
\begin{threeparttable}[b]
\begin{tabular}{cccccccc}
\hline
System & \begin{tabular}[c]{@{}c@{}}Feature\\ Extraction\end{tabular} & \begin{tabular}[c]{@{}c@{}}Decision\\ Method\end{tabular} & \begin{tabular}[c]{@{}c@{}}Unknown\\ Authentication\end{tabular} & Dataset & \#Subjects & \begin{tabular}[c]{@{}c@{}}Accuracy\\ (\%)\end{tabular} & Platform \\ \hline
\multirow{3}{*}{Tantawi et al. \cite{tantawi2015wavelet}} & \multirow{3}{*}{\begin{tabular}[c]{@{}c@{}}Discrete\\ Wavelet\\ Transform\end{tabular}} & \multirow{3}{*}{\begin{tabular}[c]{@{}c@{}}Neural\\ Network\end{tabular}} & \multirow{3}{*}{No} & \multirow{3}{*}{PTB} & \multirow{3}{*}{290} & \multirow{3}{*}{97.7} & \multirow{3}{*}{PC} \\
 &  &  &  &  &  &  &  \\
 &  &  &  &  &  &  &  \\
\multirow{3}{*}{Hammad et al. \cite{hammad2018multimodal}} & \multirow{3}{*}{CNN} & \multirow{3}{*}{\begin{tabular}[c]{@{}c@{}}Support\\ Vector\\ Machine\end{tabular}} & \multirow{3}{*}{No} & \multirow{3}{*}{PTB} & \multirow{3}{*}{290} & \multirow{3}{*}{98.66} & \multirow{3}{*}{PC} \\
 &  &  &  &  &  &  &  \\
 &  &  &  &  &  &  &  \\
\multirow{3}{*}{Thentu et al. \cite{thentu2021ecg}} & \multirow{3}{*}{CNN} & \multirow{3}{*}{\begin{tabular}[c]{@{}c@{}}Neural\\ Network\end{tabular}} & \multirow{3}{*}{No} & \multirow{3}{*}{PTB} & \multirow{3}{*}{290} & \multirow{3}{*}{97.47} & \multirow{3}{*}{PC} \\
 &  &  &  &  &  &  &  \\
 &  &  &  &  &  &  &  \\
\multirow{3}{*}{Prakash et al. \cite{prakash2022baed}} & \multirow{3}{*}{CNN} & \multirow{3}{*}{\begin{tabular}[c]{@{}c@{}}Neural\\ Network\end{tabular}} & \multirow{3}{*}{No} & \multirow{3}{*}{\begin{tabular}[c]{@{}c@{}}PTB\\ ECGID\end{tabular}} & \multirow{3}{*}{\begin{tabular}[c]{@{}c@{}}290\\ 90\end{tabular}} & \multirow{3}{*}{\begin{tabular}[c]{@{}c@{}}99.62\\ 99.49\end{tabular}} & \multirow{3}{*}{PC} \\
 &  &  &  &  &  &  &  \\
 &  &  &  &  &  &  &  \\
\multirow{3}{*}{Sepahvand et al. \cite{sepahvand2021novel}} & \multirow{3}{*}{CNN} & \multirow{3}{*}{Distance} & \multirow{3}{*}{Yes} & \multirow{3}{*}{PTB} & \multirow{3}{*}{290} & \multirow{3}{*}{99} & \multirow{3}{*}{PC} \\
 &  &  &  &  &  &  &  \\
 &  &  &  &  &  &  &  \\
\multirow{3}{*}{Hazratifard et al. \cite{hazratifard2023ensemble}} & \multirow{3}{*}{CNN} & \multirow{3}{*}{Distance} & \multirow{3}{*}{Yes} & \multirow{3}{*}{\begin{tabular}[c]{@{}c@{}}PTB\\ ECGID\end{tabular}} & \multirow{3}{*}{\begin{tabular}[c]{@{}c@{}}290\\ 90\end{tabular}} & \multirow{3}{*}{\begin{tabular}[c]{@{}c@{}}96.8\\ 93.6\end{tabular}} & \multirow{3}{*}{PC} \\
 &  &  &  &  &  &  &  \\
 &  &  &  &  &  &  &  \\
\multirow{3}{*}{Proposed (Baseline)} & \multirow{3}{*}{CNN} & \multirow{3}{*}{Distance} & \multirow{3}{*}{Yes} & PTB & 290 & 99.15 & \multirow{3}{*}{PC} \\
 &  &  &  & MIT\tnote{*} & 47 & 98.67 &  \\
 &  &  &  & ECGID\tnote{*} & 90 & 98.77 &  \\
\multirow{3}{*}{Proposed (Optimized)} & \multirow{3}{*}{CNN} & \multirow{3}{*}{Distance} & \multirow{3}{*}{Yes} & PTB & 290 & 98.67 & \multirow{3}{*}{\begin{tabular}[c]{@{}c@{}}Embedded\\ System\end{tabular}} \\
 &  &  &  & MIT\tnote{*} & 47 & 97.96 &  \\
 &  &  &  & ECGID\tnote{*} & 90 & 98.25 &  \\ \hline
\end{tabular}
\begin{tablenotes}
\item[*] Unseen Datasets
\end{tablenotes}
\end{threeparttable}%
}
\end{table*}

For the ARM Cortex-M4 processor, CPU clock cycles of various operations are as given:- Multiplication with $3$ cycles; inversion with $1$ cycle; bits-shift with $1$ cycle; and addition with $1$ cycle \cite{arm2013arm}.
\textit{Figure \ref{16_optim_time}} shows the relative run time (in terms of CPU cycles) of various optimizations (INT8 quantization and sparsity of 20\%) against the baseline system. It demonstrates quantization and pruning optimizations achieve a lower time complexity at the expense of accuracy.

Quantization precision influences accuracy and execution time. \textit{Figure \ref{17_quan_cont}} describes the performance with different quantization precisions. {We set the $16$-bit quantization model with a 100\% CPU cycle.} From $16$-bit quantization to $4$-bit quantization, the decrease in precision reduces the authentication accuracy {(99.06\% - 98.27\%)} and time complexity {(100\% - 55\%)}. This trend has been discussed because bits-shift and addition cost fewer CPU cycles. However, there is an increase in CPU cycles from full precision (without quantization) to $16$-bit quantization {(95\% - 100\%)} because the number of additions and bits-shifts is much greater than the number of multiplications. For balancing the tradeoff between accuracy (reduction 0.09\% of PTBDB, 0.71\% of MITDB and 0.52\% of ECGIDDB) and time consumption (reduction of 26.3\%), we suggest using the INT8 format ($n=8$) to replace full precision weights {because the accuracy decreases more gently}.

The sparsity level controls the amount of pruning. A higher sparsity level corresponds to more pruning. Theoretically, more pruning will cause more information to be lost, reducing accuracy. On the other hand, pruning removes redundant calculations, reducing time complexity. The experiment verifies this fact. \textit{Figure \ref{18_prun_cont}} indicates the performance with different pruning sparsity. {We set the unpruned model with a 100\% CPU cycle.} $20\%$ sparsity is an acceptable choice because it reduces CPU cycles ($20\%$) while causing minimal accuracy degradation ($0.15\%$ for PTBDB, $0.38\%$ for MITDB, and $0.14\%$ for ECGIDDB).

\subsubsection{Power Consumption}
We estimated the average current consumption for one authentication step in nRF52 SoC and found it to be $142.8\ \mu A $ while using INT8 quantization and $20\%$ sparsity. The current consumption of a typical wearable device is assumed to be $5.8\ mA$ \cite{wong2019integrated}. Assuming a battery of $100\ mAH$, when implemented in an embedded sensor, the proposed authentication technique causes a minimal $2.4\%$ battery life reduction.

\subsubsection{Limitations and Future Work}
In testing scenarios, there is a risk of favourably biased performance when segments used for training and testing originate from the same record. Addressing this bias is an important area for future research, and implementing measures to mitigate it is essential. 
Considering the significant variability in ECG signals caused by factors such as physical activity, emotional state, and electrode placement, which may impact the system's accuracy and reliability, addressing this common issue of ECG authentication is critical.
In addition, to achieve fully self-supervised training, we constructed positive and negative pairs in terms of records rather than subjects, which means that different records from the same subjects are treated as negative pairs when they should be positive pairs. Although large-scale datasets can mitigate the negative effects of this setup, finding a more rational and efficient method of constructing positive and negative pairs remains a worthwhile future research endeavour.
Future work includes using different encoder architectures to improve generalizability, larger data sets for unsupervised pre-training, and transfer learning exploration. 
In addition, enhancing the peak detection quality may improve the performance of R2R and P2T, which is worth looking into. 

\subsection{Comparison with State of the Art}
\textit{Table \ref{system_comparision}} shows the comparison between the proposed system and state-of-the-art. The proposed system achieves high accuracy and is more generalisable than the other works mentioned in \textit{Table \ref{system_comparision}}. \cite{tantawi2015wavelet, hammad2018multimodal, thentu2021ecg} have reported performance with $290$ subjects on the PTBDB, which is lower than ours. Moreover, their system is designed as a classifier to make identification decisions that cannot be used for unknown users. \cite{prakash2022baed} reported performance on PTBDB and ECGIDDB higher than ours, but their work used Dense Layer outputs results directly rather than compared with a pre-stored database, which cannot be used for unknown users. Their method carried out an identification task. In addition, with the same authentication task, \cite{sepahvand2021novel, hazratifard2023ensemble} used CNN feature extraction and distance as the decision method. Still, they also reported a low accuracy or tested with smaller subject numbers. Compared with other frameworks, the presented system was evaluated with more comprehensive samples and two different databases to test generalization ability. More importantly, the proposed system indeed implements authentication rather than identification. Compared to the research of the same type, the proposed system achieves higher accuracy than CNN-based feature extraction systems \cite{sepahvand2021novel, hazratifard2023ensemble}. We hypothesise that the coherence between the pre-training and downstream task objectives is the primary driver behind the performance enhancement. This echoes patterns seen in other domains of deep learning, notably in computer vision, where pre-training on large datasets commonly leads to improved task performance on smaller datasets.
Our proposed framework enables deployment on embedded systems, which other research under similar tasks have not implemented. In contrast, experiments on embedded platforms utilize datasets of the same number, achieving high accuracy, low complexity, and low power consumption.

\section{Conclusion}\label{sec: concl}
Our study introduced a CNN-based ECG biometric authentication system. Multiple data preprocessing methods and training frameworks are proposed to improve the system, resulting in high accuracy and generalisability. Under the consideration of actual conditions and resource constraints, this system is optimized by quantization and pruning. Experiments show that the proposed optimization approach reduced the algorithm's complexity and energy consumption. The proposed framework preserves privacy by storing encoded values in the authentication database rather than the original ECG. 

\bibliographystyle{IEEEtran}
\bibliography{reference}

\end{document}